\documentclass[prd,nofootinbib,preprint,superscriptaddress,twocolumn,10pt]{revtex4}
\pdfoutput=1
\usepackage[utf8]{inputenc}
\usepackage{amsmath,amssymb}
\usepackage{epsfig}
\usepackage{comment}
\usepackage{bbm}
\usepackage{graphicx}
\usepackage[usenames,dvipsnames]{color}
\usepackage{float}
\usepackage{bbold}
\usepackage[colorlinks,citecolor=blue]{hyperref}
\usepackage[normalem]{ulem}

\usepackage{color}
\usepackage{subfigure}

\providecommand{\upmu}{\mu}
\providecommand{\upgamma}{\gamma}
\begin{document}
\title{Leptogenesis and neutrino mass with one right-handed neutrino and Higgs inflaton}


\author{Disha Bandyopadhyay}
\email{b.disha@iitg.ac.in}
\affiliation{Department of Physics, Indian Institute of Technology Guwahati, Assam 781039, India}

\author{Debasish Borah}
\email{dborah@iitg.ac.in}
\affiliation{Department of Physics, Indian Institute of Technology Guwahati, Assam 781039, India}
\affiliation{Pittsburgh Particle Physics, Astrophysics, and Cosmology Center, Department of Physics and Astronomy, University of Pittsburgh, Pittsburgh, PA 15260, USA}

\author{Suruj Jyoti Das }
\email{surujjd@gmail.com}
\affiliation{Particle Theory  and Cosmology Group, Center for Theoretical Physics of the Universe,
Institute for Basic Science (IBS),
 Daejeon, 34126, Korea}

\author{Nobuchika Okada}
\email{okadan@ua.edu}
\affiliation{Department of Physics, University of Alabama, Tuscaloosa, Alabama 35487, USA}

\preprint{CTPU-PTC-25-32}

\begin{abstract}
We propose a novel and minimal setup where the observed baryon asymmetry of the Universe and neutrino oscillation data can be satisfied with only one right-handed neutrino (RHN) and a second Higgs doublet with the latter being also responsible for driving cosmic inflation. While inflation is realised via non-minimal coupling of the Higgs to gravity, baryon asymmetry is generated via Affleck-Dine leptogenesis. Due to the presence of only two new fields beyond the standard model (BSM), the proposed setup remains very predictive with only a small allowed parameter space consistent with the PLANCK 2018 and ACT 2025 data simultaneously. The preferred mass spectrum of the BSM particles also keeps the detection prospects alive at terrestrial experiments.

\end{abstract}

\maketitle

\section{Introduction}
The visible matter content in the present Universe is highly asymmetric, referred to as the baryon asymmetry of the Universe (BAU), a longstanding puzzle in particle physics and cosmology \cite{Zyla:2020zbs, Planck:2018vyg}. With the standard model (SM) of particle physics falling short of providing an explanation, baryogenesis \cite{Weinberg:1979bt, Kolb:1979qa} as well as leptogenesis \cite{Fukugita:1986hr} have been the most widely studied beyond standard model (BSM) frameworks explaining the observed BAU dynamically. In leptogenesis, a non-zero lepton asymmetry is generated first which later gets converted into baryon asymmetry by electroweak sphalerons \cite{Kuzmin:1985mm}. One interesting aspect of leptogenesis is its natural realisation within canonical seesaw mechanisms like type-I \cite{Minkowski:1977sc, GellMann:1980vs, Mohapatra:1979ia, Schechter:1980gr, Schechter:1981cv}, type-II \cite{Mohapatra:1980yp, Schechter:1981cv, Wetterich:1981bx, Lazarides:1980nt, Brahmachari:1997cq} and type-III \cite{Foot:1988aq} seesaw proposed to explain non-zero neutrino mass and mixing, another observed phenomena which the SM fails to address. Leptogenesis in such seesaw models typically involve the out-of-equilibrium CP violating decay of a heavy particle like heavy right handed neutrino (RHN) in type-I seesaw. In order to generate non-zero CP asymmetry as well as to explain neutrino data, at least two copies of such RHNs are required. Additionally, with hierarchical RHN spectrum, there exists a lower bound on the scale of leptogenesis, known as the Davidson-Ibarra bound $M_1 \gtrsim 10^9$ GeV \cite{Davidson:2002qv}. While this keeps canonical leptogenesis models out of reach from terrestrial experiments, it is possible to bring the scale of leptogenesis to TeV scale via resonant enhancement with tiny mass splitting between RHNs, known as the resonant leptogenesis \cite{Pilaftsis:2003gt}.

In addition to providing evidence for baryon asymmetry of the Universe, the cosmic microwave background (CMB) observations also suggest that we live in a Universe which is homogeneous and isotropic on large scales up to an impressive accuracy \cite{Planck:2018vyg, Planck:2018jri}. Such observations, however, lead to the so called horizon and flatness problems which remain unexplained in the description of standard cosmology. The theory of cosmic inflation which introduces a phase of rapid accelerated expansion in the very early Universe is the leading paradigm solving these problems in standard cosmology \cite{Guth:1980zm,Starobinsky:1980te,Linde:1981mu}. Inflation not only solves these problems, but also makes unique predictions which can be tested at present and future CMB observations.

In this work, we propose a novel scenario where leptogenesis and neutrino mass can be realised in a minimal setup with only one RHN and an additional Higgs doublet playing the role of inflaton. The same Higgs doublet also generates the desired lepton asymmetry via the Affleck-Dine (AD) mechanism \cite{Affleck:1984fy}. With just two BSM fields namely, a RHN and a second Higgs doublet, we show that the model can satisfy the requirements of generating the correct baryon asymmetry, neutrino mass and mixing while being consistent with inflationary cosmology simultaneously\footnote{A similar setup but with type-II seesaw was proposed recently by \cite{Barrie:2021mwi, Barrie:2022cub} with follow-up studies in \cite{Kaladharan:2024bop}. Affleck-Dine Dirac leptogenesis was studied in \cite{Barrie:2024yhj} with two Higgs doublets and one massless right-handed neutrino without explaining the origin of neutrino mass.}. Combination of the two Higgs doublets helps in realising the AD leptogenesis and cosmic inflation via non-minimal coupling to gravity. In addition to constraining the model from neutrino data, PLANCK 2018 constraints on inflation \cite{Planck:2018jri} and baryon asymmetry, we also scrutinize it in view of the recently reported Atacama Cosmology Telescope (ACT) data \cite{ACT:2025fju}. While PLANCK 2018 constraints keep a large part of the parameter space allowed from the requirements of neutrino mass, successful leptogenesis and cosmic inflation, ACT 2025 data disfavors most of the parameter space leaving a small allowed region. Interestingly, constraints from cosmology and neutrino data keep the BSM degrees of freedom around the TeV corner. This allows additional probe of the model at terrestrial experiments.

This paper is organised as follows. In section \ref{sec:2}, we briefly discuss the model followed by the details of inflationary dynamics in section \ref{sec:3}. We then elaborate upon the origin of baryon asymmetry via Affleck-Dine mechanism in section \ref{sec:4}. Finally, in section \ref{sec:5}, we summarise our results and conclude.

\section{The Model}
\label{sec:2}
We consider a seesaw model with one gauge singlet right-handed neutrino $N$ which can give rise to one active neutrino mass via type-I seesaw \cite{Minkowski:1977sc, GellMann:1980vs, Mohapatra:1979ia, Schechter:1980gr, Schechter:1981cv,Yanagida:1979as, Mohapatra:1980yp}. A doublet scalar field $\eta$ playing the role of inflaton can give rise to another active neutrino mass via radiative seesaw. The scalar potential has a $Z_2$ symmetry preventing $\eta$ from acquiring a vacuum expectation value (VEV), provided the condition  \begin{equation*}
     m_\eta^2 + \dfrac{1}{2} (\lambda_3 +\lambda_4) \, v^2 > 0.
 \end{equation*}
 is satisfied. We assume this to hold in our setup. The Yukawa Lagrangian, however, breaks the $Z_2$ symmetry. The relevant part of the Yukawa Lagrangian is given by 
\begin{equation}\label{IRHYukawa}
-{\cal L} \ \supset \  Y_{\alpha} \, \bar{L}_\alpha \tilde{\eta} N  +  y_{\alpha} \, \bar{L}_\alpha \tilde{\Phi} N  +\frac{1}{2} M_N \overline{N^c}N + {\rm h.c.}
\end{equation}
where $L, \Phi$ denote the lepton and SM Higgs doublet respectively. The $Z_2$ symmetry of the scalar potential can be retained in the Yukawa Lagrangian involving $\eta$ by assigning both $\eta$ and right-handed neutrino N to be $Z_2$-odd, while all SM fields remain $Z_2$-even. With this assignment, the Yukawa interaction $ Y_{\alpha} \, \bar{L}_\alpha \tilde{\eta} N $ is $Z_2$-even and therefore does not break the symmetry. As a result, $Z_2$-violating linear or trilinear $\eta$ terms are not generated at one-loop from these couplings. However, the additional Yukawa interaction $ y_{\alpha} \, \bar{L}_\alpha \tilde{\Phi} N$ involving the SM-like Higgs doublet $\Phi$ and SM leptons introduces a mild $Z_2$ breaking. This interaction can induce terms odd in $\eta$ at one loop, but its magnitude is suppressed by the small type-I seesaw Yukawa couplings. Consequently, the induced $\eta$ VEV after electroweak symmetry breaking remains negligibly small and does not affect the phenomenology discussed in the paper. Moreover, since the reheating temperature in our scenario is above the electroweak scale, the $\eta$ field has already decayed before electroweak symmetry breaking, making the impact of any such loop-induced VEV-dependent decays insignificant. We also check the possibility of $m^2_\eta$ turning negative at some energy scales due to large Yukawa coupling with $N$. Using the one-loop renormalization-group evolution (RGE) of the scalar mass parameters we find that $\eta$ mass-squared term becomes negative only in regimes where the right-handed neutrino mass is much larger than $m_\eta$. Because of our consideration $m_\eta\gtrsim M_N$, the $\eta$ mass squared term remains positive at all energy scales, in agreement with Ref. \cite{Merle:2015gea}. A complete treatment of such $Z_2$-breaking effects via RGE corrections \cite{Oredsson:2018yho} is beyond the scope of this present work and is left for future studies. The $U(1)_L$ or global lepton number charges of the BSM fields  $\eta$ and $N$ are assigned as $-1$ and $0$ respectively. Note that the 2nd term in the above Lagrangian violates lepton number, but it does not contribute to CP asymmetry from $N$ decay. The $Z_2$-symmetric scalar potential is
\begin{align}
   V (\eta, \Phi) &=   m_{\eta}^2 |\eta|^2+\frac{\lambda_2}{2} |\eta|^4+ \mu^2_\phi |\Phi |^2 + \frac{\lambda_1}{2} |\Phi |^4 \nonumber \\
   & +\lambda_3 (\eta^\dagger \eta) (\Phi^\dagger \Phi) + \lambda_4 (\Phi^\dagger \eta)(\eta^\dagger \Phi) \nonumber \\
   & + \frac{\lambda_5}{2} [ (\Phi^\dagger \eta)^2 +{\rm h.c.} ],\label{eq:Pot1}
\end{align}
where $\Phi$ is the SM Higgs doublet. The $Z_2$-symmetric scalar potential keeps the two Higgs doublets in the alignment limit \cite{Bernon:2015qea, Bernon:2015wef}. The scalar fields $\Phi$ and $\eta$ can be parameterized as
\begin{align}
    \Phi=\frac{1}{\sqrt{2}} \begin{pmatrix}
        0 \\
        \phi +v
    \end{pmatrix}, \eta= \begin{pmatrix}
        \eta^\pm \\
        \frac{(H^0 +i A^0)}{\sqrt{2}}
    \end{pmatrix}.
\end{align}
When the neutral component of $\Phi$ acquires a non-zero VEV $v$, one of the light neutrinos acquire mass via type-I seesaw
\begin{equation}
    (m^{I}_\nu)_{\alpha \beta} = -\frac{y_\alpha y_\beta v^2}{2 M_N}.
\end{equation}
Another light neutrino acquires mass at one-loop level with $\eta, N$ in the loop. The corresponding mass is given by
\begin{align}
(m^{\rm loop}_{\nu})_{\alpha \beta} \ =  \frac{Y_{\alpha}Y_{\beta} M_N}{32 \pi^2} \left[L_k(m^2_{H^0})-L_k(m^2_{A^0})\right] \, ,
\label{numass1}
\end{align}
where the function $L_k(m^2)$ is defined as 
\begin{align}
L_k(m^2) \ = \ \frac{m^2}{m^2-M^2_k} \: \text{ln} \frac{m^2}{M^2_k} \, .
\label{eq:Lk}
\end{align}
Such a neutrino mass model with a combination of tree level and radiative contributions arising from one RHN and two Higgs doublets was discussed originally by Grimus and Neufeld \cite{Grimus:1989pu}. Unlike the minimal type-I seesaw with two RHNs, this model can also explain the neutrino mass hierarchy naturally, without fine-tuning \cite{Ibarra:2011gn}.

In order to incorporate the constraints from light neutrino masses, we use the 
Casas-Ibarra (CI) parametrisation \cite{Casas:2001sr} for type-I seesaw in combination with the one for scotogenic model \cite{Toma:2013zsa}, as done for scoto-seesaw scenarions in \cite{Leite:2023gzl, Borah:2025hpo}. This is given by 
\begin{equation}
    \mathcal{Y} = \sqrt{X^{-1}_M} R \sqrt{m^{\rm diag}_\nu} U^\dagger
\end{equation}
where $R$ is an arbitrary complex orthogonal matrix satisfying $RR^{T}=\mathbbm{1}$ and $U$ is the usual Pontecorvo-Maki-Nakagawa-Sakata (PMNS) mixing matrix which diagonalises the light neutrino mass matrix in a basis where charged lepton mass matrix is diagonal. The combined Dirac Yukawa coupling $\mathcal{Y}$ is 
\begin{equation}
    \mathcal{Y} = \begin{pmatrix}
        Y^{1\times 3}_{\alpha} \\
    y^{1 \times 3}_{\alpha}
    \end{pmatrix}
    \label{eq:yuk}
\end{equation}
and $X_M$ is given by
\begin{align}
   X_M &= \begin{pmatrix}
        -X_1 & 0 \\
        0 & X_0
    \end{pmatrix}, \,\, X_1 = \frac{M_{N}}{32 \pi^2} \bigg ( L_k(m^2_{H^0})-L_k(m^2_{A^0}) \bigg ), \nonumber \\
    X_0 & =\frac{v^2}{2M_{N}}.
\end{align}
The $R$ matrix for 2 heavy neutrino scenario is given by \cite{Ibarra:2003up}
\begin{equation}
    R = \begin{pmatrix}
        0 & \cos{z_1} & \pm \sin{z_1} \\
        0 & -\sin{z_1} & \pm \cos{z_1}
    \end{pmatrix}
\end{equation}
where $z_1=a+ib$ is a complex angle. The diagonal light neutrino mass matrix, assuming normal hierarchy (NH), is given by 
\begin{equation}
    m^{\rm diag}_\nu = \begin{pmatrix}
        0 & 0 & 0 \\
        0 & \sqrt{\Delta m^2_{\rm sol}} & 0 \\
        0 & 0 & \sqrt{\Delta m^2_{\rm atm}+\Delta m^2_{\rm sol}}
    \end{pmatrix}.
\end{equation}
For inverted hierarchy (IH), $m^{\rm diag}_\nu $ is
\begin{equation}
    m^{\rm diag}_\nu = \begin{pmatrix}
        \sqrt{ \Delta m^2_{\rm atm}-\Delta m^2_{\rm sol} } & 0 & 0 \\
        0 & \sqrt{ \Delta m^2_{\rm atm} } & 0 \\
        0 & 0 & 0
    \end{pmatrix}.
\end{equation}
The PMNS mixing parameters and mass squared differences are obtained from \cite{Zyla:2020zbs}.


\section{Affleck-Dine Inflationary Dynamics}
\label{sec:3}
As mentioned earlier, we consider the Affleck-Dine (AD) mechanism \cite{Affleck:1984fy} to generate the visible sector asymmetry. We adopt a simple non-supersymmetric setup where baryogenesis occurs via leptogenesis. In a typical AD mechanism of this type, a lepton number (L) carrying field, to be referred to as the AD field hereafter, breaks L explicitly by virtue of its quadratic term. The cosmological evolution of the AD field then leads to the generation of lepton asymmetry. In the scalar potential given in Eq. \eqref{eq:Pot1}, the $\lambda_5$ term can be identified as the explicit $L$-breaking term relevant for generation of lepton asymmetry.

In order to generate asymmetry through the Affleck-Dine mechanism, the $U(1)_L$  breaking $\lambda_5$ term in the potential given by Eq. \eqref{eq:Pot1} should be effective during the inflationary and reheating dynamics. Inflation in such a case can be realised  by a combination of the inert doublet $\eta$ and the SM Higgs $\Phi$ \cite{Lebedev:2011aq, Barrie:2024yhj}, with non-minimal couplings to gravity of the form \cite{Bezrukov:2007ep}
\begin{align}
    f(|\Phi|,|\eta|)= \xi_1 |\Phi|^2 +  \xi_2 |\eta|^2\,. 
\end{align}
Here, $\xi_{1,2}$ are dimensionless couplings of the respective scalar doublets to gravity. Earlier works on a common origin of inflation and baryogenesis or leptogenesis via AD mechanism can be found in \cite{Charng:2008ke, Hertzberg:2013jba, Hertzberg:2013mba, Takeda:2014eoa, Babichev:2018sia, Cline:2019fxx, Cline:2020mdt, Lin:2020lmr, Lloyd-Stubbs:2020sed, Kawasaki:2020xyf, Mohapatra:2021aig, Barrie:2021mwi, Mohapatra:2022ngo, Borah:2022qln, Lloyd-Stubbs:2022wmh, Mohapatra:2025bdl}.
Considering $\Phi \equiv  \begin{pmatrix}
        0 \\
    \frac{1}{\sqrt{2}} \rho_1 e^{i \theta_1}
    \end{pmatrix}$ and $\eta \equiv \begin{pmatrix}
        0 \\
    \frac{1}{\sqrt{2}} \rho_2 e^{i \theta_2}
    \end{pmatrix}$, in order to realise the inflationary trajectory in the large field limit, following the analysis of Ref. \cite{Lebedev:2011aq, Barrie:2022cub, Barrie:2024yhj}  we have 
\begin{align}
    \frac{\rho_1}{\rho_2}\equiv \text{tan} \alpha = \sqrt{\frac{\lambda_2 \xi_1 -(\lambda_3 + \lambda_4)\xi_2}{\lambda_1 \xi_2 -(\lambda_3 + \lambda_4)\xi_1}} \simeq \sqrt{\frac{\lambda_2 \xi_1}{\lambda_1 \xi_2 }}\, ,
\end{align}
where, from now on,  we consider $\lambda_3, \lambda_4$ to be small enough for simplicity, without any loss of generality. The inflation can now be described in terms of a single field $\varphi$, such that
\begin{align}
    \rho_1 = \varphi~\text{sin}~ \alpha\,,\,\,\,\, \rho_2 = \varphi~\text{cos}~ \alpha\,,\,\,\,\, \xi = \xi_1\text{sin}^2~ \alpha + \xi_2\text{cos}^2~ \alpha\,.
    \nonumber 
\end{align}
The potential can then be written as\footnote{While radiative corrections to the potential can be important, we can always tune the scalar couplings at low energy scale appropriately to get the desired values at the scale of inflation after incorporating one-loop corrections.}
\begin{align}
   V(\varphi, \theta)  = \frac{m^2}{2}\varphi^2 + \frac{1}{4}\lambda \varphi^4  + 2 \tilde \lambda_5 \cos 2\theta \varphi^4 ~,
\end{align}
with $\lambda=\frac12(\lambda_1\sin^4\alpha+\lambda_2\cos^4\alpha)$, $\tilde \lambda_5=\frac{\lambda_5}{8} \sin^2\alpha\cos^2\alpha$, $\theta = \theta_2 - \theta_1$ and $m^2=m_{\eta}^2 \cos^2 \alpha+ \mu_{\phi}^2 \sin^2 \alpha$.


Going to the Einstein frame through the conformal transformation
\begin{align}
    \tilde{g}_{\mu \nu} = \Omega^2g_{\mu \nu}~,~~~ \Omega^2= 1+\xi\varphi^2/M_P^2\,, 
\end{align}
results in the following Einstein frame Lagrangian
\begin{align}
    \frac{\mathcal L}{\sqrt{-g}} & =-\frac{M_P^2}{2}  R - \frac{1}{2}  g^{\mu\nu} \partial_\mu \chi\partial_\nu \chi  -
\frac{1}{2} f(\chi)   g^{\mu\nu} \partial_\mu \theta \partial_\nu\theta \nonumber \\
& - U(\chi,\theta)\,\label{eqn:Lag_eins}
\end{align}
where
\begin{align}
     f(\chi) \equiv   \frac{\varphi(\chi)^2 \cos^2 \alpha}{\Omega^2(\chi)}~,  ~~ U(\chi,\theta)  \equiv   \frac{V(\varphi(\chi), \theta)}{\Omega^4(\chi)}~,\label{eq:f}
\end{align}
and $R$ represents the Ricci scalar. The canonical scalar field $\chi$ is obtained through the field redefinition given by
\begin{align}
    \frac{d \chi}{d \varphi} = \frac{\sqrt{6 \xi^2 \varphi^2/M_P^2 + \Omega^2}}{\Omega^2}. \label{eq:fieldtrans}  
\end{align}
In the large field limit, the potential reduces to that of the well-known Starobinsky inflation, given by
\begin{align}
    U_{\text{inf}}(\chi)= \frac{3}{4}\Lambda^2 M_P^2 \left(1-e^{-\sqrt{\frac{2}{3}}\frac{\chi}{M_P}}\right)^2~
\end{align}
where $\Lambda=\sqrt{\frac{\lambda M_P^2}{3\xi^2}}$ and $M_P$ is the reduced Planck mass. 

\noindent \textbf{Inflationary predictions:} Here, we briefly review the predictions for the inflationary observables of Starobinsky inflation, using the slow-roll formalism. The slow- roll parameters are defined as
\begin{align}
    \epsilon_V \equiv \frac{M_{\rm P}^2}{2} \left(\frac{V^{\prime}}{V}\right)^2, \quad
    \eta_V \equiv M_{\rm P}^2\, \frac{V^{\prime \prime}}{V}  \,, \label{eq:sr}
\end{align}
where $'$ indicates derivative w.r.t. the conformal field $\chi$. At the end of inflation, we have $\epsilon_V(\chi_{\rm end})=1$, which gives
\begin{align}
    \chi_{\rm end}=  M_{\rm P} \sqrt{\frac{3}{2}} \ln\left(1+\frac{2}{\sqrt{3}}\right)\,.
\end{align}
Now, the total number of e-folds starting from the epoch when the CMB pivot scale ($\sim 0.05 ~\text{Mpc}^{-1}$) exit the horizon during inflation till the end of inflation can be written as
\begin{align}
    N_* = \int^{\chi_\star}_{\chi_{\text{end}}} \frac{1}{\sqrt{2\, \epsilon_V(\chi)}}\, \frac{d\chi}{M_{\rm P}} \,,
\end{align}
where $\chi_*$ indicates the field value at the horizon exit. With the above quantities in hand, the inflationary observables can be  calculated as

\noindent (i) Amplitude of scalar perturbation:  $A_s = \frac{1}{24\pi^2\, \epsilon_V}\, \frac{V}{M_{P}^4}\, $, \\
(ii) Scalar spectral index: $ n_s = 1- 6\epsilon_V + 2\eta_V\, $, \\
(iii) Tensor-to-scalar ratio: $r =16 \epsilon_V \,$. \\

The measured value of $A_s= 2.1 \times 10^{-9}$ \cite{Planck:2018jri} fixes $\Lambda$. The parameters $N_*, r, \chi_*,  \Lambda$ can be expressed as functions of $n_{s}$ as follows \cite{Drees:2025ngb}:
\begin{align}
    N_*&\simeq \frac{2}{1-n_s}+\frac34 \ln(1-n_s)+0.47+\mathcal{O}(1-n_s)\,,\label{eq:N_*}\\
    r&\simeq 3(1-n_s)^2 + \frac{9}{2}(1-n_s)^3 + \mathcal{O}[(1-n_s)^4]\,,\label{eq:r}\\
    \chi_* &\simeq \sqrt{\frac32} M_P \ln{\left(\frac{7+4\sqrt{4-3n_s}-3n_s}{3(1-n_s)}\right)}\,, \label{eq:chi*}\\
\Lambda& \simeq \pi\, M_{\rm P} \sqrt{\frac{A_s}{6}} \left(1 + 2\sqrt{4 - 3\,n_s} - 3\,n_s\right).\label{eq:Lmda}
    \end{align}
    
\noindent {\bf Reheating dynamics and constraints:}
The relation between the fields $\varphi$ and $\chi$ given by Eq. \eqref{eq:fieldtrans}, and consequently the potential in Eq. \eqref{eq:f}  take different forms during inflation and the reheating stage after inflation. Depending on the field values, we have distinct regimes which determine the background dynamics of the Universe. From Eq. \eqref{eq:fieldtrans}, we get
\begin{widetext}
\begin{equation}
 \dfrac{\chi}{M_P}
 \approx
 \left\{
 \begin{array}{lll}
 \dfrac{\varphi}{M_P}
 & \mbox{for $\dfrac{\varphi}{M_P} \ll\dfrac{1}{\xi}$} 
 \\
 \sqrt{\dfrac{3}{2}}\,\xi\left(\dfrac{\varphi}{M_P}\right)^2  & \mbox{for $\dfrac{1}{\xi}\ll \dfrac{\varphi}{M_P} \ll \dfrac{1}{\sqrt{\xi}}$\quad} 
 \\
 \sqrt{\dfrac{3}{2}} \ln\Omega^2 =
 \sqrt{\dfrac{3}{2}} \ln\left[1+\xi\left(\dfrac{\varphi}{M_P}\right)^2\right] \quad
 & \mbox{for $\dfrac{1}{\sqrt{\xi}}\ll \dfrac{\varphi}{M_P}$} 
 \end{array}
 \right.
 \label{chi-varphi-relation_sm1}
 \end{equation}
 \end{widetext}
The potential during these regimes can be approximated as
\begin{widetext}
\begin{equation}
U(\chi) \approx
\left\{
\begin{array}{lll}
\dfrac{1}{2}m^{2}\chi^2 
& \mbox{for $\dfrac{\chi}{M_P}\ll \sqrt{\dfrac{2}{\lambda}} \dfrac{m}{M_{P}}$}
& \mbox{(matter-like 2)}
\\
\dfrac{1}{4}\lambda\chi^4 
& \mbox{for   $ \sqrt{\dfrac{2}{\lambda}} \dfrac{m}{M_{P}}\ll\dfrac{\chi}{M_P}\ll\dfrac{1}{\xi}$}
& \mbox{(radiation-like)}
\\
\dfrac{1}{2} \Lambda^2 \chi^2 
& \mbox{for $\dfrac{1}{\xi} \ll \dfrac{\chi}{M_P} \ll 1$\quad}
& \mbox{(matter-like 1)}
\\
\dfrac{3}{4}\Lambda M_P^2
\left(1-e^{-\sqrt{\frac{2}{3}}(\chi/M_P)}\right)^2 \quad
& \mbox{for $1\ll \dfrac{\chi}{M_P}$}
& \mbox{(inflation)}
\end{array}
\right.
\label{Uchi_sm1}
\end{equation}
\end{widetext}
Since we are interested in perturbative reheating, we consider $\xi\simeq 30$, to avoid the non-perturbative regime \cite{Sfakianakis:2018lzf}. In such a case, the equation of state parameter $w$ changes from $w=-1$ to $w=1/3$ after a few e-folds (to be determined explicitly later), passing quickly through the matter-like 1 era given by the $3^{\rm rd}$ line in Eq. \eqref{Uchi_sm1}. We determine the effect of this changing equation of state numerically, which will be explained shortly. Subsequently, the potential becomes dominated by the quartic term giving rise to the radiation-dominated era.  As the magnitude of the field $\chi$ decreases further, at some point the quadratic term starts dominating, and we have a matter-like era with $w=0$.


The reheating temperature can be related to the inflationary observables, particularly $n_s$, which places severe constraints depending on the equation of state $w$ during reheating. After the end of inflation, the Universe evolves from  $w\simeq -1$ to a radiation background with $w=1/3$. We can write the e-folding number from the end of inflation to end of reheating as
\begin{widetext}
\begin{align}
    N_{\rm re}=\ln{\left(\frac{a_{\rm re}}{a_{\rm end}}\right)}&\simeq \ln{\left(\frac{a_{\rm rd}}{a_{\rm end}}\frac{a_{\rm eq}}{a_{\rm rd}}\frac{a_{\rm re}}{a_{\rm eq}}\right)}= \ln{\left(\frac{a_{\rm rd}}{a_{\rm end}} \right)}+\frac{1}{3(1+w_{\rm 1})}\ln{\left(\frac{\rho_{\rm rd}}{\rho_{\rm eq}}\right)}+ \frac{1}{3(1+w_{\rm 2})}\ln{\left(\frac{\rho_{\rm eq}}{\rho_{\rm re}}\right)}\,\nonumber\\ &= \mathcal{C}+\frac{1}{3(1+w_{\rm 1})}\ln{\left(\frac{\rho_{\rm rd}}{\rho_{\rm eq}}\right)}+ \frac{1}{3(1+w_{\rm 2})}\ln{\left(\frac{\rho_{\rm eq}}{\rho_{\rm re}}\right)}.\label{eq:Nre}
\end{align}
\end{widetext}
In the above, $a_{\rm end}, a_{\rm rd},a_{\rm eq}, a_{\rm re}$\footnote{Throughout this paper, we use the following notation for
subscripts: 
``end'' denotes quantities evaluated at the end of inflation; 
``rd'' refers to the beginning of the radiation-dominated phase (with \(w = 1/3\)); 
``eq'' indicates the matter like 2–radiation equality point;
and ``re'' corresponds to the end of reheating.} denotes the scale factor at the end of inflation, beginning of radiation dominated potential with $w=1/3$, matter like 2 -radiation equality and end of reheating respectively, while $\rho$ indicates the corresponding energy densities. To be more precise, we include the factor $\mathcal{C}=\ln{\left(\frac{a_{\rm rd}}{a_{\rm end}}\right)}$ from our numerical calculations, which takes into account the effect of the changing equation of state from $w=-1$ to $w=1/3$, as discussed earlier. Considering $\xi=30$ as explained above and fixing $\lambda$ from the scalar power spectrum, we find $\mathcal{C}\simeq4.2$.  Additionally, the total energy density at the beginning of radiation-dominated era is determined numerically as $\rho_{\rm rd}\simeq 8 \times 10^{-18} M_P^4$. $a_{\rm eq}$ indicates the  scale factor when the quadratic term starts to dominate the potential energy with energy density $\rho_{\rm eq}=2 m^4/\lambda$. Also, in our case, we have $w_1=w_{\rm RD}= 1/3, ~w_2=w_{\rm MD}=  0 $, but in the following, we keep $w_1,w_2$ to make the results more general. 

Now, since at horizon exit we have $k= a_k H_k$ with $a_k, H_k$ being the corresponding scale factor and Hubble parameter respectively, we can write
\begin{align}
    &\ln{\left(\frac{k}{a_k H_k}\right)}= \ln{\left[\frac{k}{a_0 H_k}\frac{a_{\rm end}}{a_k}\frac{a_{\rm re}}{a_{\rm end}}\frac{a_0}{a_{\rm re}}\right]}=0\nonumber\\
    \implies &\ln{\left(\frac{k}{a_0 H_k}\right)}+ N_*+ N_{\rm re} + \ln{\left(\frac{a_0}{a_{\rm re}}\right)}=0\,,\label{eq:kaH}
\end{align}
where $a_0$ denotes the present value of the scale factor. $H_k$ indicates the Hubble at horizon-crossing which can be determined using the definition of $r= \frac{P_h}{P_{\zeta}}$, where $P_{h}= \frac{2 H^2}{\pi^2 M_P^2}$ is the amplitude of tensor perturbations, while $P_{\zeta}$ represents the corresponding amplitude of scalar perturbations. Thus, we have $H_k= \frac{\pi M_P \sqrt{A_s r}}{\sqrt{2}}$.

Using conservation of entropy from the end of reheating until the present temperature, we have
\begin{align}
   a_{\rm re}^3 g_{*s} T_{\rm RH}^3= a_0^3\left(2 T_0^3 + 6 \frac78 T_{\rm \nu 0}^3\right)\,,
\end{align}
where $T_{\rm RH}$ is the reheating temperature, $T_0= 2.73$ K is the present temperature and $T_{\nu 0}=\left(\frac{4}{11}\right)^{1/3}T_0$ indicates the temperature of neutrinos. Hence, we can write
\begin{align}
    \frac{a_{\rm re}}{a_0}= \left(\frac{43}{11 g_{*s}}\right)^{1/3} \frac{T_0}{T_{\rm RH}} = \left(\frac{43}{11 g_{*s}}\right)^{1/3}\left(\frac{\pi^2 g_*T_0^4}{30 \rho_{\rm re}}\right)^{1/4}\,.\label{eq:ara0}
\end{align}
Using the above in Eq. \eqref{eq:Nre}, we get
 \begin{align}
    \rho_{\rm re} =  \exp{\left(-3 N'_{\rm re}(1+w_2)\right)} \left(\frac{\rho_{\rm rd}}{\rho_{\rm eq}}\right)^{\frac{1+w_2}{1+w_1}} \rho_{\rm eq}\,,\label{eq:rhoreh}
 \end{align}
where we define $N'_{\rm re}= N_{\rm re}- \mathcal{C}$. Note that the above expression reduces to the one with a single equation of state \cite{Cook:2015vqa, Drewes:2017fmn, Ghoshal:2022fud, Becker:2023tvd, Zharov:2025evb, Biswas:2025adi}, in the limit $w_1=w_2$. Eq. \eqref{eq:ara0} can be rewritten as
\begin{align}
    \ln{\frac{a_{\rm re}}{a_0}} & = \frac13\ln{\left(\frac{43}{11 g_{*s}}\right)} +\frac14 \ln{\left(\frac{\pi^2 g_*}{30}\right)}+\frac34 N'_{\rm re}(1+w_2) \nonumber \\
    & + \frac14 \ln{\left(\frac{\rho_{\rm eq}}{ \rho_{\rm rd}}\right)^{\frac{1+w_2}{1+w_1}}}+\frac14 \ln{\left(\frac{T_0^4}{\rho_{\rm eq}}\right)}\,.
\end{align}
Using the above expression in Eq. \eqref{eq:kaH}, we get
\begin{align}
    N'_{\rm re} &=\frac{4}{3 w_2-1}\bigg [N_*+\frac13 \ln{\left(\frac{11 g_{*s}}{43}\right)}+\frac14\ln{\left(\frac{30}{\pi^2g_*}\right)} \nonumber \\
    & +\ln{\left(\frac{k}{a_0 T_0}\right)}-\frac14\ln{\left(\frac{\rho_{\rm eq}}{ \rho_{\rm rd}}\right)^{\frac{1+w_2}{1+w_1}}} \nonumber\\ 
     & -\frac14 \ln{\left(\frac{\pi^4M_P^4A_s^2r^2}{4\rho_{\rm eq}}\right)}+\mathcal{C}\bigg]\,.\label{eq:Nre2} 
\end{align}
Note again, that in the limit $w_1=w_2$, the above relation of $N_{\rm re}$ matches the one with a single equation of state \cite{Cook:2015vqa, Drewes:2017fmn, Ghoshal:2022fud, Becker:2023tvd, Zharov:2025evb, Biswas:2025adi}. Finally, using Eq. \eqref{eq:rhoreh}, we can express Eq. \eqref{eq:Nre2} in terms of the reheating temperature as 
\begin{align}
    T_{\rm RH}&= \rho_{\rm eq}^{1/4} \left(\frac{30}{\pi^2g_*}\right)^{\frac{1}{1-3w_2}}\left( \frac{\rho_{\rm rd}}{\rho_{\rm eq}}\right)^{\frac{1+w_2}{(1+w_1)(1-3w_2)}} \nonumber \\
    & \times \left(\frac{11g_{*s} k^3}{43 a^3_0 T^3_0}\right)^{\frac{1+w_2}{1-3w_2}} \left(\frac{\pi^4M_P^4A_s^2r^2}{4\rho_{\rm eq}}\right)^{\frac{3(1+w_2)}{4(3w_2-1)}} \nonumber \\ 
    & \times \exp{\left(\frac{3(1+w_2)}{1-3w_2}(N_*+\mathcal{C})\right)}\,.\label{eq:TRH1}
\end{align}
Here, $\rho_{\rm eq} = 2 m^4/\lambda$ and  $k/a_0 = 0.05\, \text{Mpc}^{-1}$ is the CMB pivot scale. Thus, the reheating temperature can be written in terms of $n_s$ using the expressions relating $N_*, r, \Lambda$ with $n_s$ mentioned before (cf. Eqs. \eqref{eq:N_*}- \eqref{eq:Lmda}). In our case, for $w_1=1/3$, $w_2=0$, Eq. \eqref{eq:TRH1} gives
\begin{align}
    T_{\rm RH} & =8~\rho_{\rm eq}^{1/4} \left(\frac{30}{\pi^2g_*}\right)\left(\frac14 \frac{\rho_{\rm rd}}{\pi^4M_P^4A_s^2r^2}\right)^{\frac{3}{4}} \left(\frac{11g_{*s}}{43}\right) \nonumber \\
    & \times \left(\frac{k}{a_0 T_0}\right)^{3} \exp{(3 (N_*+\mathcal{C}))}\,.\label{eq:TRH2}
\end{align}
On the other hand, considering perturbative reheating from the decay $\eta \rightarrow N L$, we can also estimate the reheat temperature as
\begin{align}
    T_{\rm RH} \approx \sqrt{M_P \Gamma_\eta}\,, \,\, \Gamma_\eta &= \frac{Y^\dagger_\alpha Y_\alpha}{8\pi} m_\eta \left ( 1-\frac{M^2_N}{m^2_\eta} \right )^2 \nonumber\\ &\sim \frac{Y^\dagger_\alpha Y_\alpha}{2\pi} m_\eta \left ( \frac{\Delta m}{m_\eta} \right )^2. 
    \label{eq:trh}
\end{align}
In the last step of the above equation, we have assumed $m_\eta + M_N \sim 2 m_\eta$ which is valid for small mass splitting $\Delta m$.
\begin{figure}[h]
\includegraphics[scale=0.6]{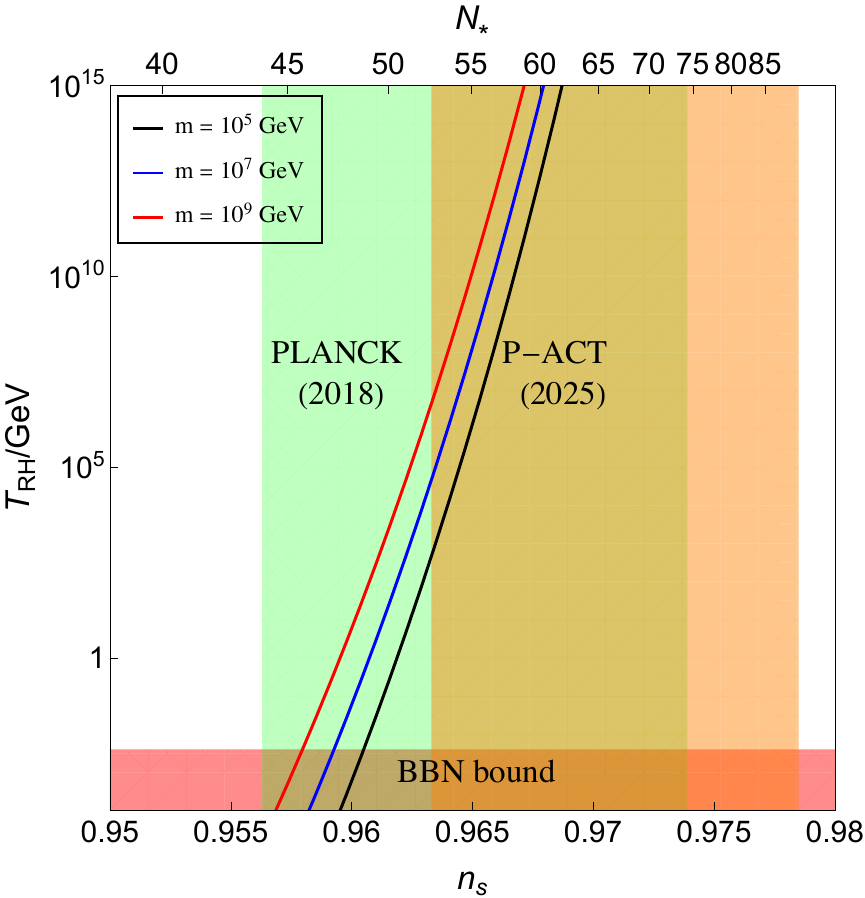} 
\caption{Variation of reheat temperature $T_{\rm RH}$ with scalar spectral index $n_s$ in our model for three different choices of $m$. The corresponding values of e-folding no. $N_*$ are denoted in the upper x-axis.}
\label{fig:nsTR}  
\end{figure}  

Fig. \ref{fig:nsTR} shows the variation of reheat temperature $T_{\rm RH}$ with $n_s$ for three different benchmark choices of $m$. The pink shaded region at the bottom is disfavoured due to the lower bound on reheat temperature $T_{\rm RH} > T_{\rm BBN} \simeq 4$ MeV \cite{Kawasaki:2000en}. The vertical shaded regions correspond to $2\sigma$ allowed values of $n_s$ from PLANCK 2018 and PLANCK-ACT (P-ACT) 2025 data. Clearly, P-ACT 2025 restricts $T_{\rm RH}$ above a certain lower bound while PLANCK 2018 data allows $T_{\rm RH}$ as low as a few MeV. In contrast to a single equation of state ($w \leq 1/3$) during the reheating era in Starobinsky inflation, which is disfavored in light of the recent P-ACT 2025 data as it predicts very high reheating temperatures \cite{Zharov:2025evb,Drees:2025ngb, Liu:2025qca, Haque:2025uis, Yogesh:2025wak}, our scenario can accommodate lower values of $T_{\rm RH}$. This is because in the presence of multiple equation of states during reheating (cf. Eq. \eqref{Uchi_sm1}), the scaling of $T_{\rm RH}$ with $n_s$ changes in a non-trivial manner (cf. Eq. \eqref{eq:TRH1}). This leads to a decrease in the slope of $T_{\rm RH}$ with $n_s$ (cf. Fig. \ref{fig:nsTR}), as compared to the case with a single $w$ (see for instance Ref. \cite{Drees:2025ngb}).

\section{Affleck-Dine leptogenesis}
\label{sec:4}
In this section, we discuss the generation of baryon asymmetry via Affleck-Dine leptogenesis. From the Lagrangian in Einstein frame given by Eq. \eqref{eqn:Lag_eins}, the equations of motion for $\chi$ and $\theta$ are derived to be \cite{Barrie:2024yhj}  
\begin{eqnarray}
&& \ddot{\chi} -\frac{1}{2} f'(\chi)  \dot \theta^2 + 3 H \dot \chi + U_{,\chi} =0  ~,\nonumber \\
&& \ddot{\theta} + \frac{f'(\chi)}{f(\chi)}  \dot \theta \dot \chi + 3 H \dot \theta +\frac{1}{f(\chi)} U_{,\theta} =0 ~.
\label{eq:thetaEOM}
\end{eqnarray} 
Here $'$ in $f$ denotes derivative w.r.t. $\chi$. During inflation, $\theta$ is in the approximate slow-roll phase, during which we have
\begin{align}
    \dot{\theta} \simeq -\frac{M_P U_{,\theta}}{f(\chi)\sqrt{3U}}.\label{eq:thetadot}
\end{align}
Because of the presence of the $U(1)_L$ breaking term, asymmetry number density $n_L\propto \varphi^2 \dot{\theta}$ is generated, which at the end of inflation is given by
\begin{align}
  \left.n_L\right|_\text{end}&= Q_L \varphi^2_\text{end} (\chi) \dot{\theta}_\text{end} \text{cos}^2\alpha\\
  &\simeq \frac{Q_L M_P U_{, \theta} \Omega^2}{\sqrt{3U_{\text{end}}}}\nonumber\\
   &\simeq \frac{8Q_L\tilde{\lambda}_5 \sin(2\theta_{\text{end}})}{\sqrt{3\lambda}\xi} M_P^3\,,\label{eq:nlend}
\end{align}
where we used Eq. \eqref{eq:f} and the slow-roll condition for $\dot{\theta}$ given by Eq. \eqref{eq:thetadot}.  In the last step, we have taken the quartic term to be dominant at the end of inflation. In Eq. \eqref{eq:nlend}, $Q_L$ denotes the $U(1)_L$ charge of the AD field $\eta$.

The final asymmetry depends on the details of the reheating dynamics. To understand this clearly, let us look into Eq. \eqref{eq:thetaEOM} for $\theta$, which can be rewritten as
\begin{align}
    \frac{d}{dt}(a^3 f(\chi)\dot{\theta})=\frac{d}{dt}\left(\frac{n_L a^3}{Q_L \Omega^2}\right)= -a^3 U_{, \theta}\,\label{eq:nL1}\,,
\end{align}
where we have substituted $\dot{\theta}$ in terms of $n_L$ and used Eq. \eqref{eq:f} for $f(\chi)$. In addition to the form of the $\chi$ potential during reheating, the asymmetry also depends on the form of the $U(1)_L$ breaking term, which is determined by $U_{, \theta}$. 
\begin{figure*}
    \centering
   \includegraphics[scale=0.35]{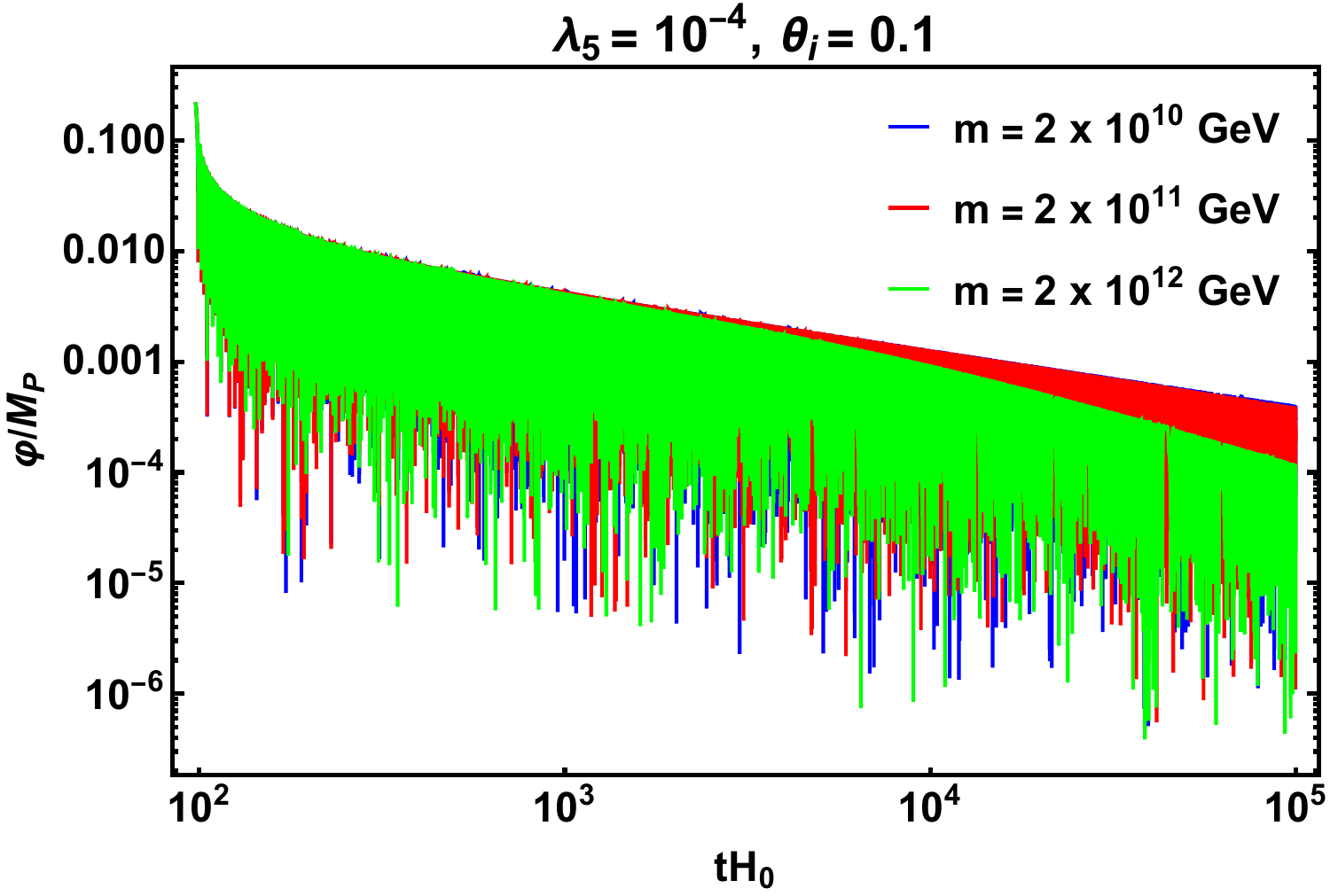}~~\includegraphics[scale=0.35]{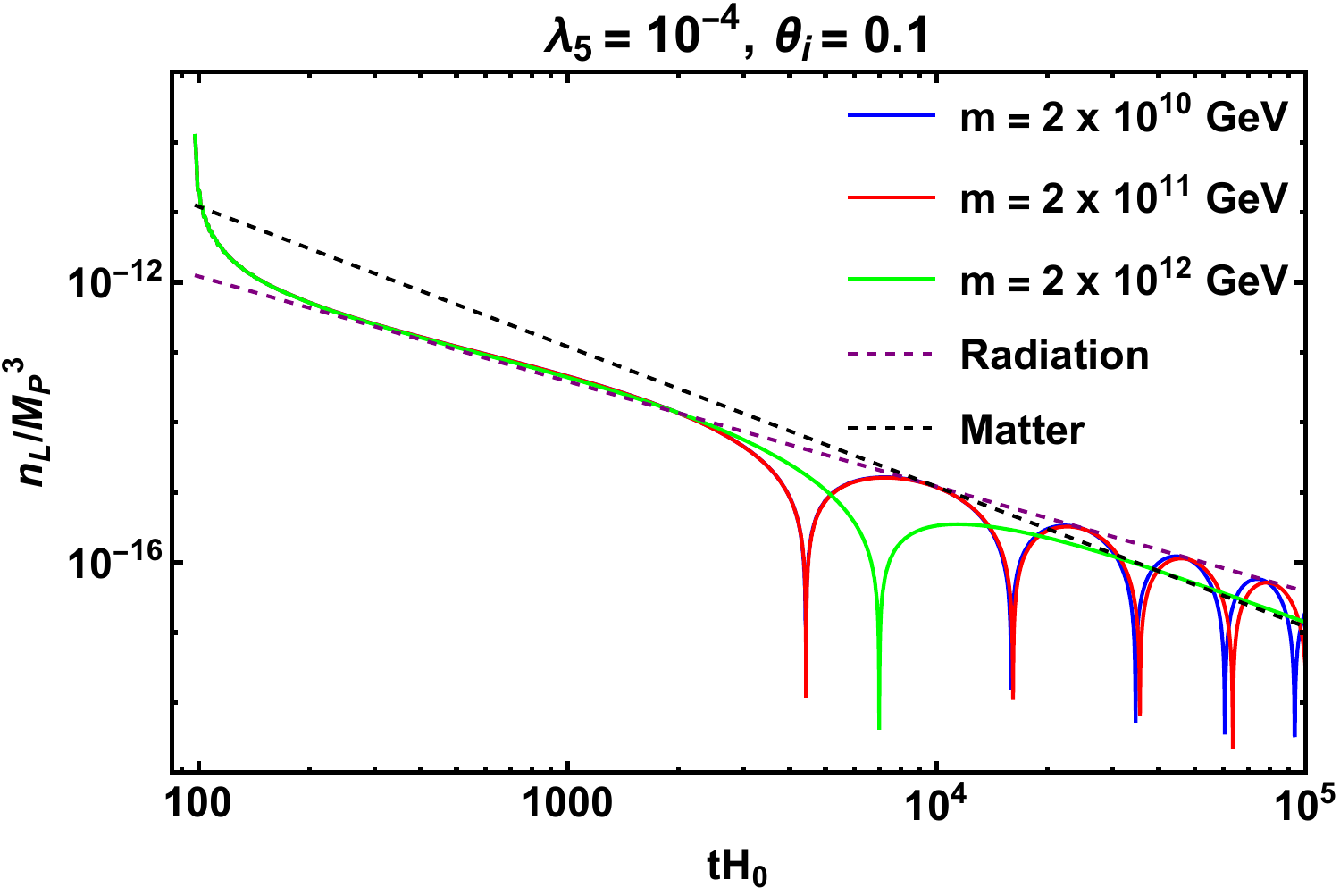} \\
      \includegraphics[scale=0.35]{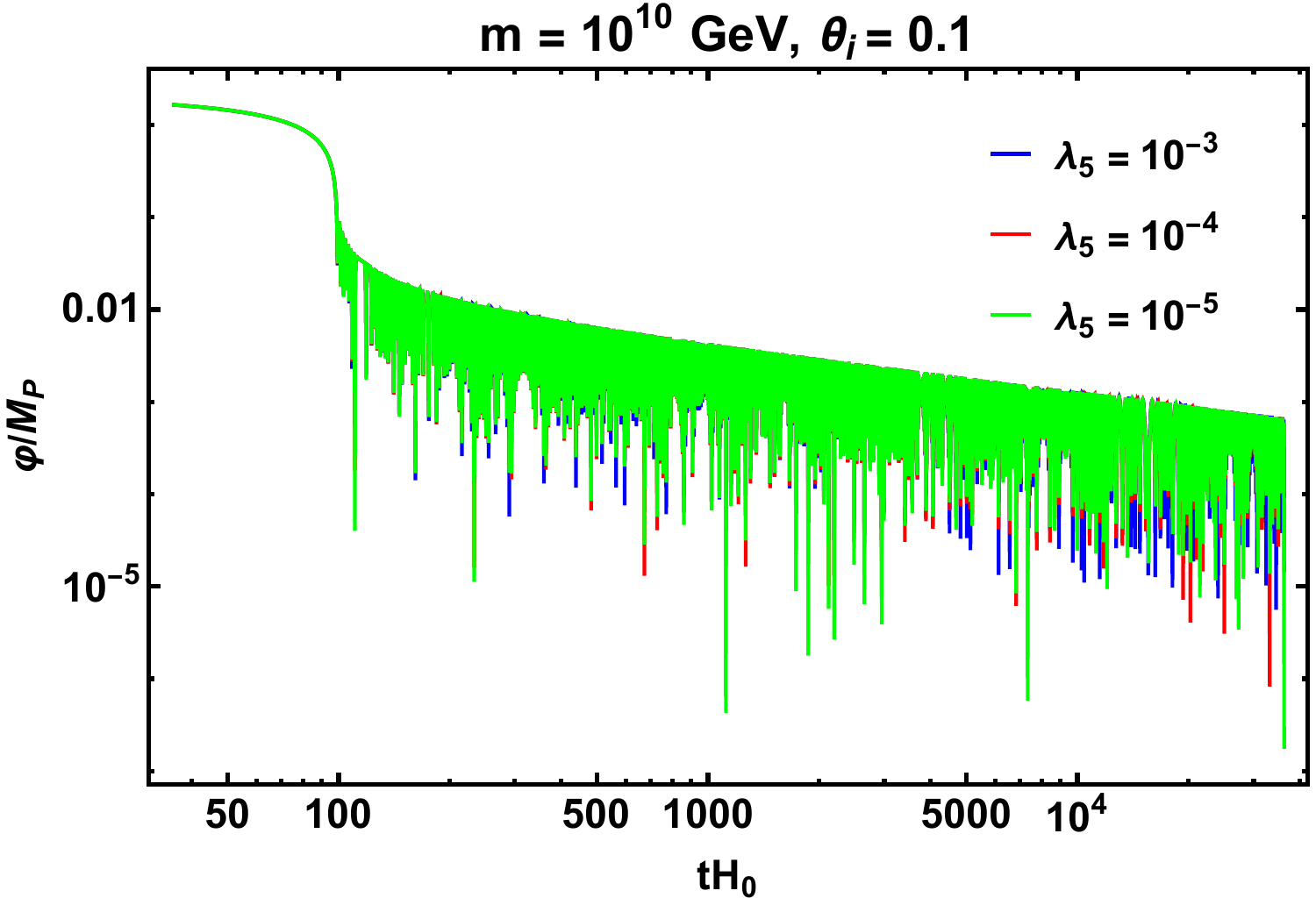}~~\includegraphics[scale=0.35]{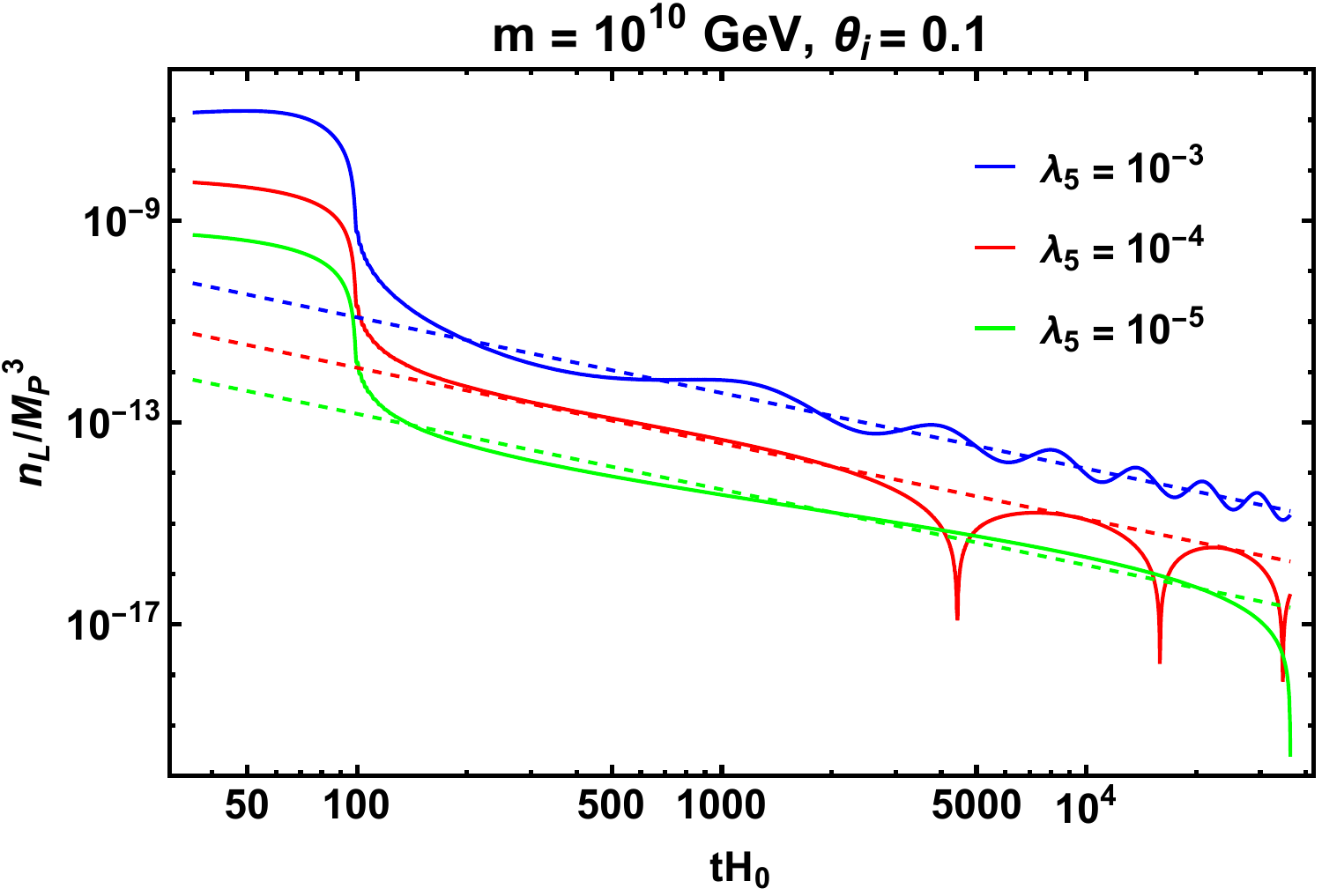} \\
        \includegraphics[scale=0.35]{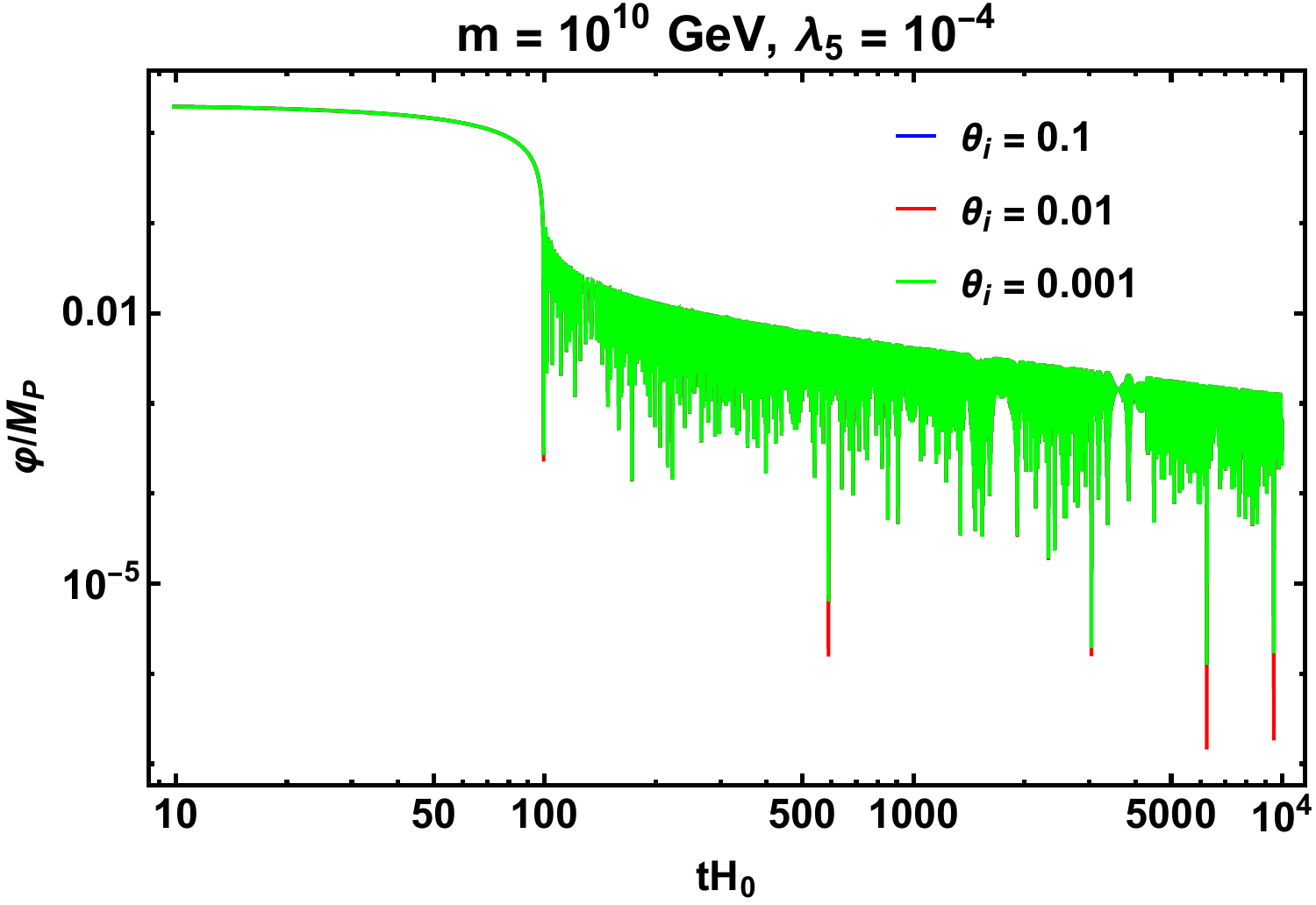}~~\includegraphics[scale=0.35]{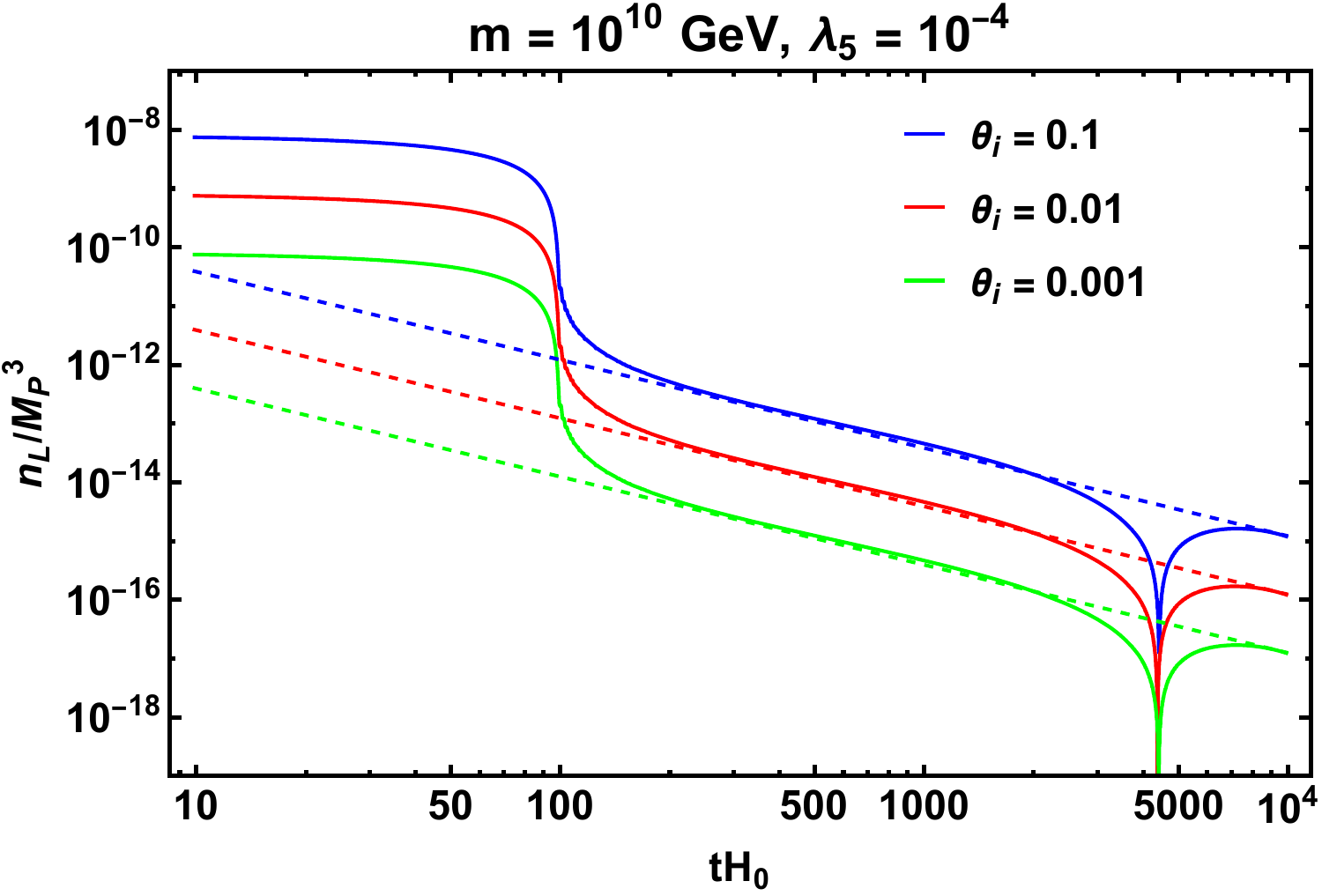}
    \caption{Evolution of $\varphi$ (left panel) and the asymmetry density $n_L\propto \varphi^2 \dot{\theta}$ (right panel) for three different values of $m$ (top panel),  $\lambda_{5}$ (middle panel), $\theta_i$ (bottom panel) and choosing $\xi_2 =30, ~\xi_1=1$. The dotted lines in the top right panel denote the scaling of the asymmetry density as matter, i.e. $n_L\propto a^{-3}\propto t^{-2}$ or radiation, i.e. $n_L\propto a^{-3}\propto t^{-3/2}$, whereas for the middle and bottom right panels the dotted lines denote the scaling of asymmetry as radiation  i.e. $n_L\propto a^{-3}\propto t^{-3/2}$.}
    \label{fig:num1}
\end{figure*}

Now, if  $U_{, \theta}$ redshifts faster than $a^{-3}$ (matter-like), then the right hand side of Eq. \eqref{eq:nL1} becomes diminishing over time and the quantity $\frac{n_L a^3}{Q_L \Omega^2}$ remains conserved. After the end of inflation, the potential during reheating is initially dominated by the quartic term after a few e-folds as discussed above. Hence $a^{3}U_{, \theta}\propto a^{3} \chi^4\propto a^{-1} $ becomes subdominant with time. When the magnitude of $\varphi= \chi$ reduces to $\sqrt{\frac{2}{\lambda}}m$, the potential energy density behaves like matter. In this case, we again have $a^{3}U_{, \theta}\propto a^{3} \chi^4 \propto a^{-3}$, which remains subdominant. Thus, the quantity   $\frac{n_L a^3}{Q_L \Omega^2}$ remains approximately conserved, after the start of radiation domination with the quartic potential.

\begin{figure*}
    \centering
   \includegraphics[scale=0.35]{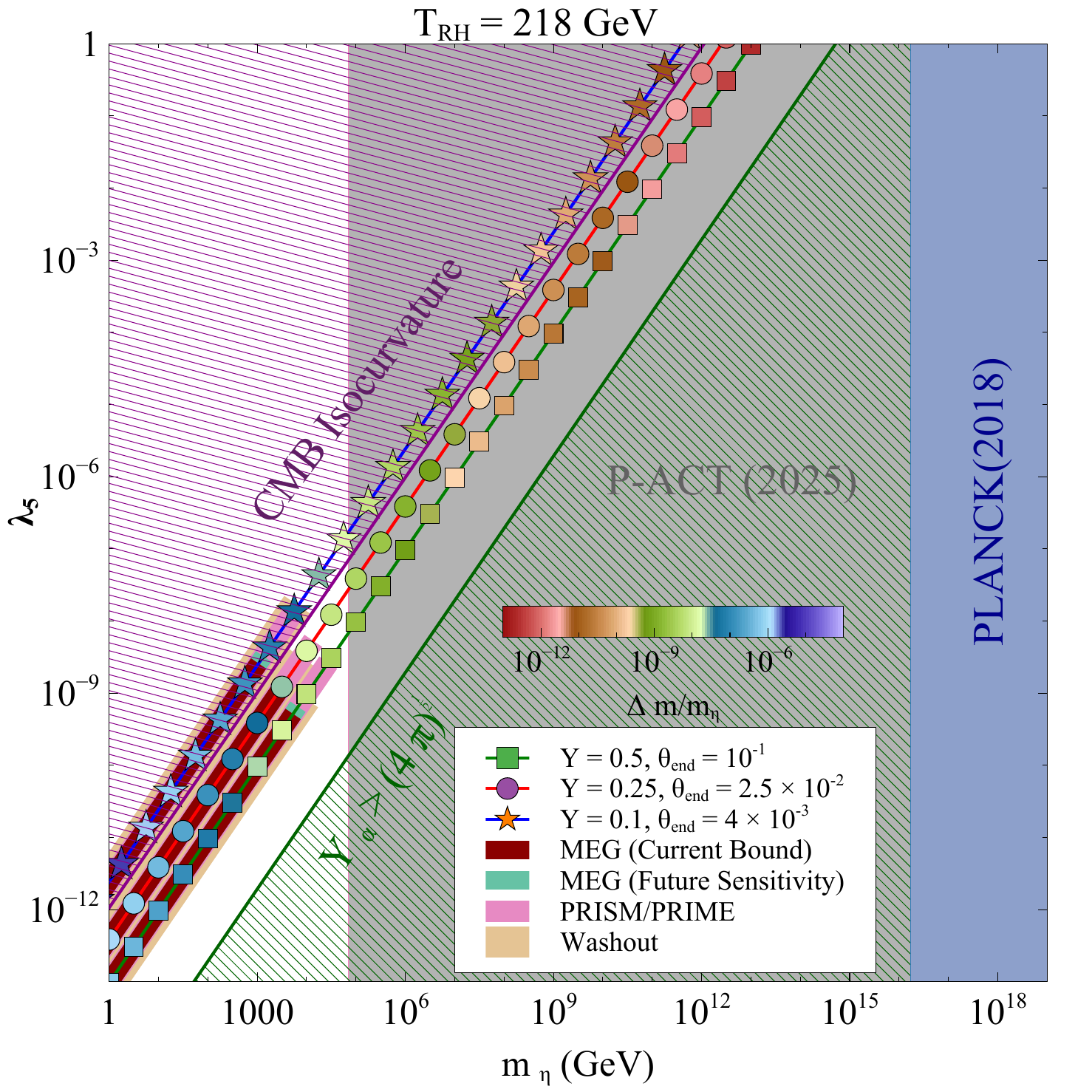}~~\includegraphics[scale=0.35]{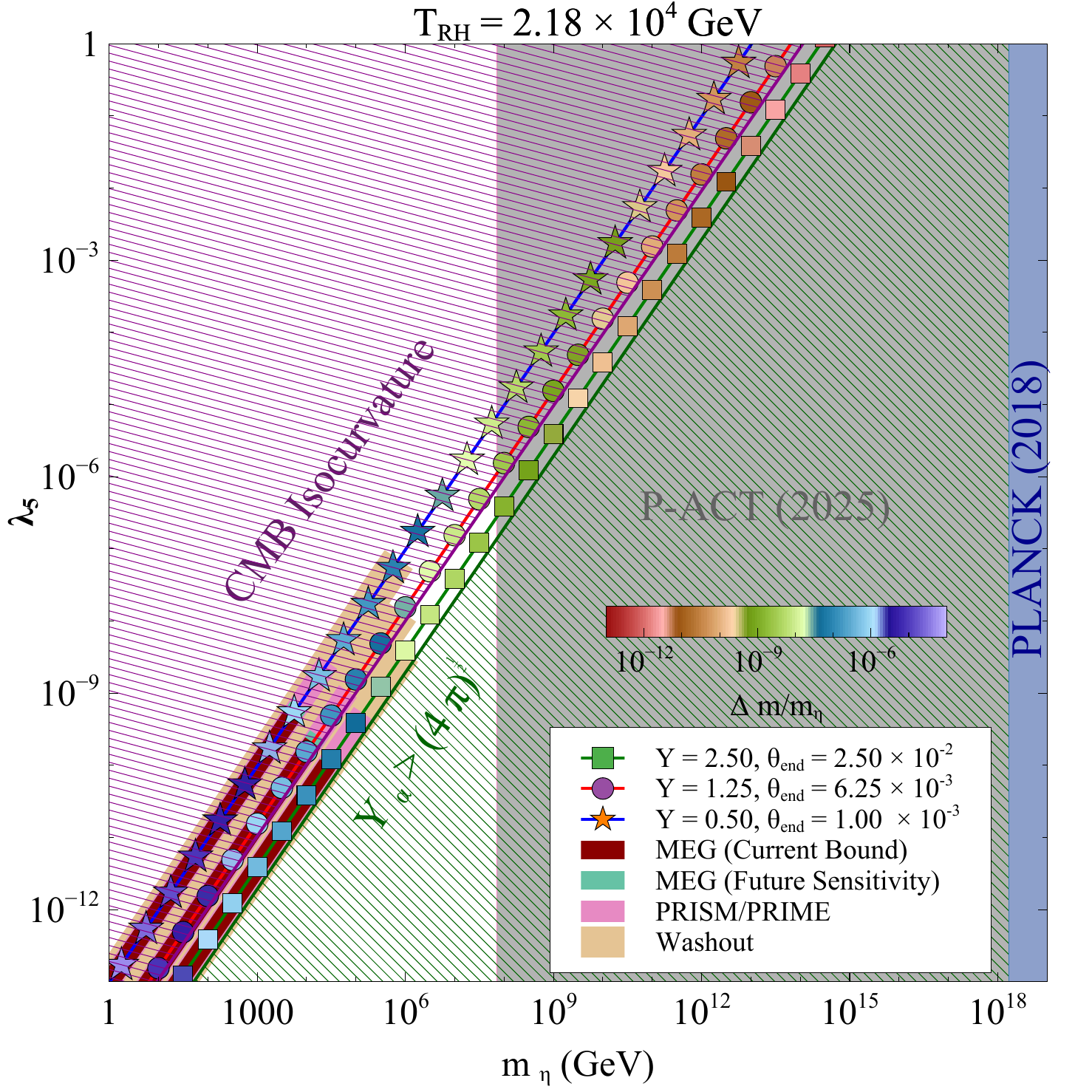}
    \caption{Allowed parameter space for successful leptogenesis in $\lambda_5-m_\eta$ plane for $z_1 = 0.2\, - i\, 0.1$, $Y = (\sum_{\alpha} Y^\dagger_\alpha Y_\alpha)^{1/2}$ and $T_{\rm RH} =$ 218 GeV (Left) and  $T_{\rm RH} = 2.18 \times 10^4$ GeV (Right). The gray and blue shaded regions are disfavored by P-ACT 2025 and PLANCK 2018 data respectively. The meshed region in green is disfavored by perturbative upper limit on Yukawa coupling. The isocurvature bound from CMB given by Eq. \eqref{isoc}, rules out the regions with too small $\theta_{\rm end}$, which is shown by the purple meshed region. The yellow shaded region corresponds to the regime of strong washout whereas the dark red band shows the parameter space disfavored by MEG II constraints on $\mu \to e \gamma$ \cite{MEGII:2025gzr}. The cyan colored band corresponds to the parameter space within future sensitivity of MEG II \cite{MEGII:2018kmf} while the pink colored band indicates future sensitivity of  PRISM/PRIME to $\mu \rightarrow e$ conversion \cite{BARLOW201144}. }
    \label{fig:res}
\end{figure*}

The lepton number density at the end of reheating can be obtained by integrating Eq. \eqref{eq:nL1}, which gives
\begin{widetext}
\begin{align}
    n_L(t_{\rm re})&\simeq \frac{\Omega^2(t_{\rm re})}{\Omega^2(t_{\rm end})}\left(\frac{a(t_{\rm end})}{a(t_{\rm re})}\right)^3\left[n_L|_\text{end} -\Omega^2(t_{\rm end})\int^{t_{\rm re}}_{t_{\rm end}}U_{,\theta} ~\frac{a^3}{a(t_{\rm end})^3} dt \right]\nonumber\\
    &\simeq  \,n_L|_\text{end} \frac{\Omega^2(t_{\rm re})}{\Omega^2(t_{\rm end})}\exp{(-3\mathcal{C})}\left(\frac{\chi(t_{\rm eq})}{\chi(t_{\rm rd})}\right)^3\left(\frac{t_{\rm eq}}{t_{\rm re}}\right)^2 \left[1+c_2 \right]\,,\label{eq:nltrh1} 
\end{align}
\end{widetext}
The constant $c_2=-\int^{t_{\rm re}}_{t_{\rm end}}\frac{\sqrt{3 U_{\rm end}}}{M_P} ~\frac{a^3}{a(t_{\rm end})^3} dt $  indicates the effect of the source term during reheating and is determined numerically.  Here, we used Eq. \eqref{eq:nlend} to write $U_{,\theta}$ in terms of $n_L|_\text{end}$. We find $c_2\simeq 55.5$. Recall that after the beginning of radiation-like behavior of the potential, its effect becomes subdominant, as explained above. $t_{\rm eq}$ denotes the time when the quadratic term  in the potential dominates and can be written as
\begin{align}
    t_{\rm eq} = t_{\rm rd} \left(\frac{\chi(t_{\rm rd})}{\chi(t_{\rm eq})}\right)^2\,.
\end{align}
Using the above in Eq. \eqref{eq:nltrh1}, we get
\begin{align}
    n_L(t_{\rm re}) & \sim \frac{1.4\,\exp{(-3\mathcal{C})}\, g_{*}\tilde{\lambda}_5\sin{2\theta}}{\xi \lambda^{1/4}}\frac{\Omega^2(t_{\rm re})}{\Omega^2(t_{\rm end})}\left(\frac{T_{\rm RH}}{m}\right) \nonumber \\
    & \times \left(\frac{T_{\rm RH}}{\rho_{\rm rd}^{1/4}}\right)^3M_P^3 ~(1+c_2)\,. \label{eq:nL}
\end{align}
where we used Eq. \eqref{eq:nlend}.  Recall that $\mathcal{C}$ and $\rho_{\rm rd}$ are obtained numerically as explained below Eq. \eqref{eq:Nre}. The final baryon number yield after the decoupling of electroweak sphalerons is obtained as 
\begin{align}\label{asymm}
    Y_{B}\sim \frac{\exp{(-3\mathcal{C})}\, \tilde{\lambda}_5\sin{2\theta_{\text{end}}}}{\xi \lambda^{1/4}}\frac{\Omega^2(t_{\rm re})}{\Omega^2(t_{\rm end})}\left(\frac{T_{\rm RH}}{m}\right)\frac{M^3_{P}}{\rho_{\rm rd}^{3/4}} ( 1+c_2)\,.
\end{align}

Lepton asymmetry can also be washed out due to $L$ violating scattering processes with interactions of the type $\eta \Phi^\dagger \leftrightarrow \eta^\dagger \Phi $ or $L \Phi (\eta) \leftrightarrow \overline{L} \Phi^\dagger (\eta^\dagger)$. The rates for such washout processes can be calculated as 
\begin{equation}
    \Gamma_1 = \Gamma (\eta \Phi^\dagger \leftrightarrow \eta^\dagger \Phi) \approx n^{\rm eq}_\eta \frac{\lambda^2_5}{T^2}, 
\end{equation}
\begin{equation}
    \Gamma_2 = \Gamma (L \Phi \leftrightarrow \overline{L} \Phi^\dagger) \approx \frac{\lvert y^\dagger_\alpha y_\alpha \rvert^2}{8\pi M^2_N} n^{\rm eq}_L, 
\end{equation}
\begin{equation}
    \Gamma_3 = \Gamma (L \eta \leftrightarrow \overline{L} \eta^\dagger) \approx \frac{\lvert Y^\dagger_\alpha Y_\alpha \rvert^2}{8\pi M^2_N} n^{\rm eq}_\eta.
    \label{gamma3}
\end{equation}
Another source of washout comes from the inverse–decay process: $N L \to \eta$. The corresponding washout rate in comparison to the Hubble rate can be approximated as
\begin{equation}
    W_{\rm ID} = \frac{1}{4} \frac{\Gamma_\eta }{H( T= m_\eta)}\left(\frac{m_\eta}{T}\right)^3 K_1\left(\frac{m_\eta}{T}\right)
\end{equation} 
where $K_1$ is the modified Bessel function of the second kind.\\
We constrain the parameter space in such a way to keep all these washout processes out-of-equilibrium.

Finally, we make a rough estimate of the isocurvature perturbations, which can be induced by the quantum fluctuations of $\theta$ \cite{Enqvist:1998pf, Lloyd-Stubbs:2022wmh}. The variance of these fluctuations is given by \cite{Mukhanov:2005sc}
\begin{align}
    \langle\delta \theta^2\rangle \simeq \left(\frac{H_k}{2 \pi}\right)^2\frac{1}{\chi_*^2}\,,
\end{align}
The isocurvature perturbations can be related to the temperature fluctuations observed in CMB as \cite{Hu:1994uz, Dodelson:2003ft}
\begin{align}
\left(\frac{\delta T}{T}\right)_{\rm iso} \simeq  -\frac25 \frac{\Omega_B}{\Omega_{\rm m}} \frac{\delta n_B}{n_B}\,.
\end{align}
Here, $\frac{\Omega_B}{\Omega_{\rm m}}\simeq 0.16$ \cite{Planck:2018vyg} is the ratio of the baryon to matter energy densities. $n_B$ is the baryon number density, which in our case can be found using Eq. \eqref{eq:nL} as
\begin{align}
   \frac{\delta n_B}{n_B} \simeq 2 \cot{(2 \theta_i)} \delta\theta\,.
\end{align}
On the other hand, the adiabatic perturbations can be written as \cite{Hu:1994uz, Dodelson:2003ft}
\begin{align}
    \left\langle \left(\frac{\delta T}{T}\right)^2 \right\rangle_{\rm adi} \simeq \frac{1}{20}\frac{H_k^2}{8 \pi^2M_P^2 \epsilon_V}\, 
\end{align}
where $\epsilon_V$ is the slow-roll parameter defined in Eq. \eqref{eq:sr}. The ratio of these two perturbations is calculated to be
\begin{align}
   \alpha_{\rm non-adi}=\frac{ \left\langle \left(\frac{\delta T}{T}\right)^2 \right\rangle_{\rm iso}}{\left\langle \left(\frac{\delta T}{T}\right)^2 \right\rangle_{\rm adi}}\simeq \frac{128 M_P^2 ~\epsilon_V}{5}\frac{\Omega_B^2}{\Omega_{\rm m}^2} \frac{\cot^2{(2\theta_i)}}{\chi_*^2}\,.
\end{align}
Using the upper bound  $\alpha_{\rm non-adi}\lesssim 2 \times 10^{-2}$ \cite{Akrami:2018odb}, we arrive at a lower bound on $\theta_i$, given by
\begin{align}
  \theta_i \gtrsim 0.15 \sqrt{3(1-n_s)^2+\frac92(1-n_s)^3} \, 
  \label{isoc}
\end{align}
where we used Eqs. \eqref{eq:r}, \eqref{eq:chi*} for our inflationary predictions in terms of $n_s$. For $n_s$ in the range $0.95$ to $0.98$, the above bound on $\theta_i$ lies in the range $0.013$ to $0.005$. As we shall see, we have sufficient allowed parameter space for these values of $\theta_i$, which could also potentially be investigated by future CMB experiments. We leave a detailed numerical calculation of these isocurvature perturbations to future works.

\section{Results and conclusion}
\label{sec:5}
In Fig. \ref{fig:num1}, we show the evolution of $\varphi$ (left panel) and the asymmetry density $n_L\propto \varphi^2 \dot{\theta}$ (right panel) for different choices of relevant parameters. The top panel shows the variation w.r.t. $m$ parametrising the mass of the inflaton. The middle panel shows the variation w.r.t. the L-violating coupling $\lambda_5$ coupling, whereas the bottom panel indicates the effect of varying the initial value of $\theta$ field namely, $\theta_i$. As we can see, $\phi$-trajectories naturally overlap as the evolutions are very similar for changes in parameters $\lambda_5 \, \rm and \, \theta_i$, and visible differences appear only when the mass is varied. In the top right panel, we observe that the initial asymmetry after the end of inflation scales like radiation for all masses. For a larger mass $m$, the scaling later transitions to matter-like behavior, since the field reaches equality (\(t_{\text{eq}}\)) sooner and the field \(\chi\) starts oscillating as matter. In contrast, smaller $m$ continues to scale like radiation over the same period. Using Eq. \eqref{asymm}, we see that the asymmetry produced at the end of inflation is directly proportional to $\lambda_5$ and $\theta_{\text{end}}$ (or $\theta_i$). However, the subsequent scaling of the asymmetry with time is independent of $\lambda_5$ and $\theta_{\text{end}}$.

Fig. \ref{fig:res} shows the summary of our results in $\lambda_5-m_\eta$ plane with $\Delta m/m_\eta \equiv (m_\eta-M_N)/m_\eta$ being shown in the color bar. The shaded regions are disfavored by P-ACT 2025 and PLANCK 2018 data respectively whereas the meshed region in green violates perturbative upper limit on Yukawa coupling. The contours of different shapes indicate leptogenesis and neutrino mass satisfying points for different choices of Yukawa coupling $Y_\alpha$ and $\theta_{\rm end}$, while keeping the reheating temperature fixed. We have used $\theta_{\text{end}}$ and $\theta_i$ interchangeably, as $\theta_i$ is approximately equal to $\theta_{\text{end}}$, especially for smaller values of $\lambda_5$ ($10^{-4}$ and below). In this regime, the value of $\theta$ remains the same up to two significant digits. In the evolution plots, we determined the asymmetry numerically by setting the initial condition $\theta_i$, whereas in the summary plots, we used the analytical expression from Eq. \eqref{asymm} and therefore substituted $\theta_{\text{end}}$ instead. From Eq. \eqref{eq:trh}, it is clear that $\Delta m/m_\eta$ should decrease with $m_\eta$ for fixed $Y_\alpha$ and $T_{\rm RH}$. This pattern is visible from both the plots shown in Fig. \ref{fig:res}. Due to the choice of a low $T_{\rm RH}$, the corresponding relative mass difference $\Delta m/m_\eta$ needs to be fine-tuned as the corresponding Yukawa coupling can not be made arbitrarily small due to the constraints from CMB isocurvature as well as light neutrino mass. Such a small mass splitting is not technically natural in our minimal setup and additional symmetries and non-minimal particle content may be required to explain such fine-tuning, which we do not pursue further in this work. Since the reheat temperature is chosen to the small and Yukawa coupling $Y_\alpha \gtrsim \mathcal{O}(0.1)$, one requires phase-space suppression in the decay width of $\eta$ as can be noticed from the color bar in Fig. \ref{fig:res}. The yellow shaded portion of these contours indicate the region where washout effects are dominant and hence lepton asymmetry is likely to get washed out. Due to large Yukawa coupling, the washout $\Gamma_3$ in Eq. \eqref{gamma3} is the most dominant one and the yellow shaded region in Fig. \ref{fig:res} corresponds to the parameter space where 
$$\frac{\Gamma_3}{H} \bigg \rvert_{T=T_{\rm RH}} >1.$$
The red shaded region is disfavored by the limits on lepton flavor violating (LFV) decay BR$(\mu\to e\gamma)$ from MEG-II~\cite{MEGII:2025gzr} while the cyan shaded region is within future sensitivity \cite{MEGII:2018kmf}, details of which can be found in appendix \ref{appen1}. While there exist other LFV processes like $\mu \rightarrow 3e$ \cite{SINDRUM:1987nra} and $\mu \to e$ conversion \cite{SINDRUMII:2006dvw}, the corresponding constraints on the parameter space remain weaker compared to the one for $\mu \to e \gamma$. However, future sensitivity of PRISM/PRIME to $\mu \rightarrow e$ conversion \cite{BARLOW201144} remains promising, as indicated by the pink shaded band. Additionally, CMB isocurvature constraint given by Eq. \eqref{isoc} disfavors regions with too small $\theta_{\rm end} \sim \theta_i$, which is indicated by the meshed purple region. Clearly, P-ACT 2025 allows only a small region of parameter space consistent with successful leptogenesis, neutrino mass and mixing. On the other hand, a large part of the parameter space remains consistent with the PLANCK 2018 data.

To conclude, we have proposed a novel and very minimal framework for explaining the baryon asymmetry of the Universe together with neutrino oscillation data by extending the standard model of particle physics with only two new fields: one right-handed neutrino and a second Higgs doublet with the latter also playing the role of cosmic inflaton. Light neutrino masses arise from a combination of type-I and radiative seesaw with the lightest neutrino mass being zero. While canonical leptogenesis from CP-violating decay is not feasible due to the chosen minimal field content, we implement the Affleck-Dine mechanism by identifying the second Higgs doublet as the AD inflaton. Future data from cosmology and terrestrial experiments will be able to shed more light into the currently allowed parameter space of the model. The model also predicts vanishingly lightest active neutrino mass which keeps the effective neutrino mass much out of reach from ongoing tritium beta decay experiments like KATRIN \cite{KATRIN:2024cdt}, while future data should be able to confirm or refute it. Additionally, near-future observation of neutrinoless double beta decay can also falsify our scenario, particularly for normal ordering of light neutrinos. While the minimal setup proposed here does not have any particle dark matter candidate, one could explore the possibility of forming primordial black hole dark matter due to Higgs fluctuations during inflation \cite{Espinosa:2017sgp}. We leave exploration of such interesting possibilities to future works.

\acknowledgments
The work of D.B. is supported by the Science and Engineering Research Board (SERB), Government of India grants MTR/2022/000575, CRG/2022/000603 and by the Fulbright-Nehru Academic and Professional Excellence Award 2024-25. D.B. also thanks Alejandro Ibarra for insightful discussions and comments on the manuscript. S.J.D. is supported by IBS under the project code IBS-R018-D1. S.J.D. also thanks Sougata Ganguly for some useful discussions. The work of N.O. is supported in part by the United States Department of Energy Grant Nos. DE-SC0012447 and DE-SC0023713.

\appendix

\section{Charged lepton flavor violation}
\label{appen1}
The same particles taking part in generating one-loop mass for one of the neutrinos can also contribute to charged lepton flavor violating decays like $\mu \rightarrow e \gamma$. A detailed discussion of charged lepton flavor violation in seesaw model with one RHN and two Higgs doublets can be found in \cite{Dudenas:2022von}. The branching ratio for $\mu \rightarrow e \gamma$ is given by
\[
\mu \to e + \gamma,
\]
where the branching ratio is given by
\[
\mathrm{Br}\bigl(\mu \to e \gamma\bigr)
= \frac{3 \,(4\pi)^3\,\alpha_{\mathrm{em}}}{4\,G_F^2}\,\bigl| A_D\bigr|^2 \, \mathrm{Br}\bigl(\mu \to e \, \nu_\mu \, \overline{\nu_e}\bigr),
\] 
with
\[
F_2(x) = \frac{1 - 6x + 3x^2 + 2x^3 - 6x^2 \log x}{6\,(1 - x)^4},
\]
and
\[
A_D
= \frac{Y_e^*Y_\mu}{2\,(4\pi)^2}\,\frac{1}{m_{\eta\pm}^2}\,F_2\Bigl(\frac{M_N^2}{m_{\eta\pm}^2}\Bigr).\\
\]
\noindent Here, \(\alpha_{\mathrm{em}}\) is the electromagnetic fine structure constant, and \(G_F\) is the Fermi constant.
The current experimental upper limit is given by
\[
\mathrm{Br}\bigl(\mu \to e \gamma\bigr) < 3.1\times 10^{-13}\,, \] from the combined analysis of MEG I and MEG II \cite{MEGII:2023ltw} with the future sensitivity being BR$(\mu\rightarrow e\gamma)<6\times10^{-14}$ \cite{MEGII:2018kmf}.


\providecommand{\href}[2]{#2}\begingroup\raggedright\endgroup

\end{document}